\documentclass[11pt]{article}
\usepackage{moriond,epsfig}

\def\Journal#1#2#3#4{{#1} {\bf #2}, #3 (#4)}

\def\NIMA{{\em Nucl. Instrum. Methods} A}

\def\PLB{{\em Phys. Lett.}  B}
\def\PRL{\em Phys. Rev. Lett.}
\def\PRD{{\em Phys. Rev.} D}

\def\be{\begin{equation}}
\def\ee{\end{equation}}
\def\bea{\begin{eqnarray}}
\def\eea{\end{eqnarray}}

\begin{document}
\vspace*{4cm}
\title{Charm Physics Results from CLEO}

\author{ {\it The CLEO Collaboration} \\ M.S. DUBROVIN }

\address{Department of Physics \& Astronomy, Wayne State University,\\
Detroit, MI 48201}

\maketitle\abstracts{
We present recent results in the charm sector
from the CLEO Collaboration. The data used were collected
by the CLEO II and III detectors at the
Cornell Electron Storage Ring (CESR).
Measurements of form factors and branching fractions
in semileptonic decays of D-meson and charmed-baryons,
searches for CP violation and rare decays, and
Dalitz analyses of D-meson decays are discussed,
as well as future plans for the CLEO-c experiment at CESR-c.
}

\section{Charm physics at CLEO}

Since the 1980's the primary goal of CLEO Collaboration was 
the study of $\Upsilon$-resonances and $B$-meson physics. 
The charm sector received steady 
attention as well. Among $\sim 350$ CLEO publications
about 18\% were devoted to the decays of charmed mesons and
about 14\% to charmed baryons. 
At this time we continue to study all aspects of charm including
semileptonic decays of mesons and baryons, Dalitz plot analysis
of three-body D-meson decays, 
fragmentation, inclusive decay rates, doubly and singly 
Cabibbo suppressed decays, 
searches for rare and forbidden decays, CP-violation, etc.

The data using in our analyses were collected with three detector 
configurations \cite{CLEOII}$^-$\cite{CLEOIII},
\[ 
\begin{array}{lcl}
     CLEO~II   & Oct.1989 - Apr.1995 & 4.7~fb^{-1},      \\
     CLEO~II.V & Nov.1995 - Feb.1999 & 9~fb^{-1},        \\
     CLEO~III  & Jul.2000 - Mar.2003 & \sim 16~fb^{-1}, 
\end{array}
\]
at Cornell $e^+e^-$ Storage Ring (CESR) with $\sqrt{s} \simeq 10~GeV$.
At this time about half of the CLEO~III data is available for analysis. 

\section{Recent results overview}


\subsection{First Observation of the Exclusive Decays
             $\Lambda_c^+ \to \Lambda \pi^+\pi^+ \pi^- \pi^0$
         and $\Lambda_c^+ \to \Lambda \omega \pi^+$}

\begin{figure}[!htb]
  \begin{minipage}[t]{8cm}
    \vspace{-2cm}    
    \leftline{\rotatebox[x=20mm,y=-20mm]{90}{\bf Events}}
    \vspace{-2cm}
    \hspace*{5mm} \includegraphics*[width=8cm,bb=34 253 490 480]{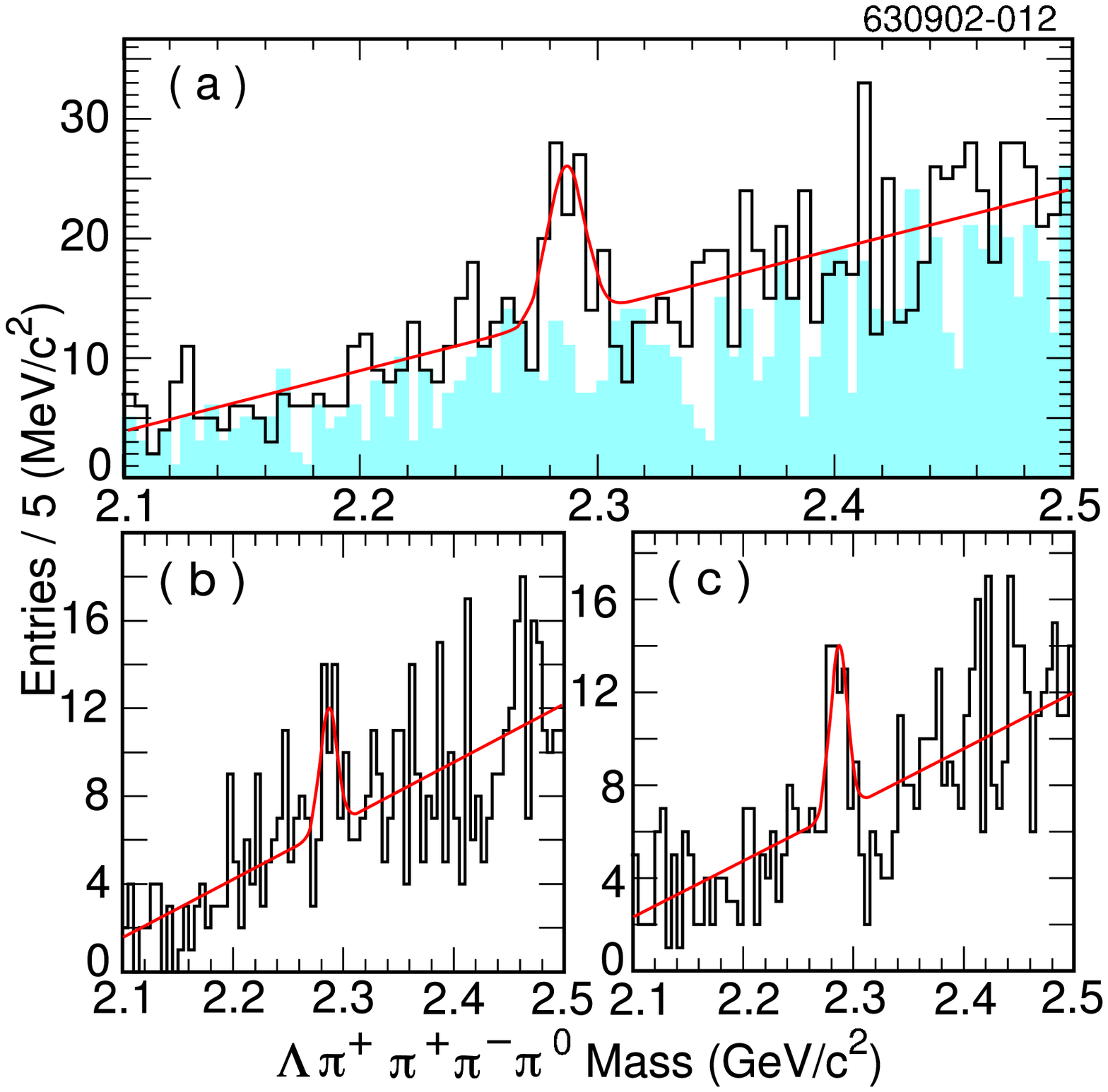}
    \hspace*{10mm}{ \Large $ \Lambda \pi^+\pi^+ \pi^- \pi^0$ Mass (GeV/c$^2$)}
    \caption{\label{fig:blusk1} Invariant $\Lambda \pi^+\pi^+ \pi^- \pi^0$ mass spectrum.}
  \end{minipage}
 \hfill
  \begin{minipage}[t]{6cm}
   \vspace*{1mm}
    \epsfxsize=6cm
    \rightline{\epsfbox{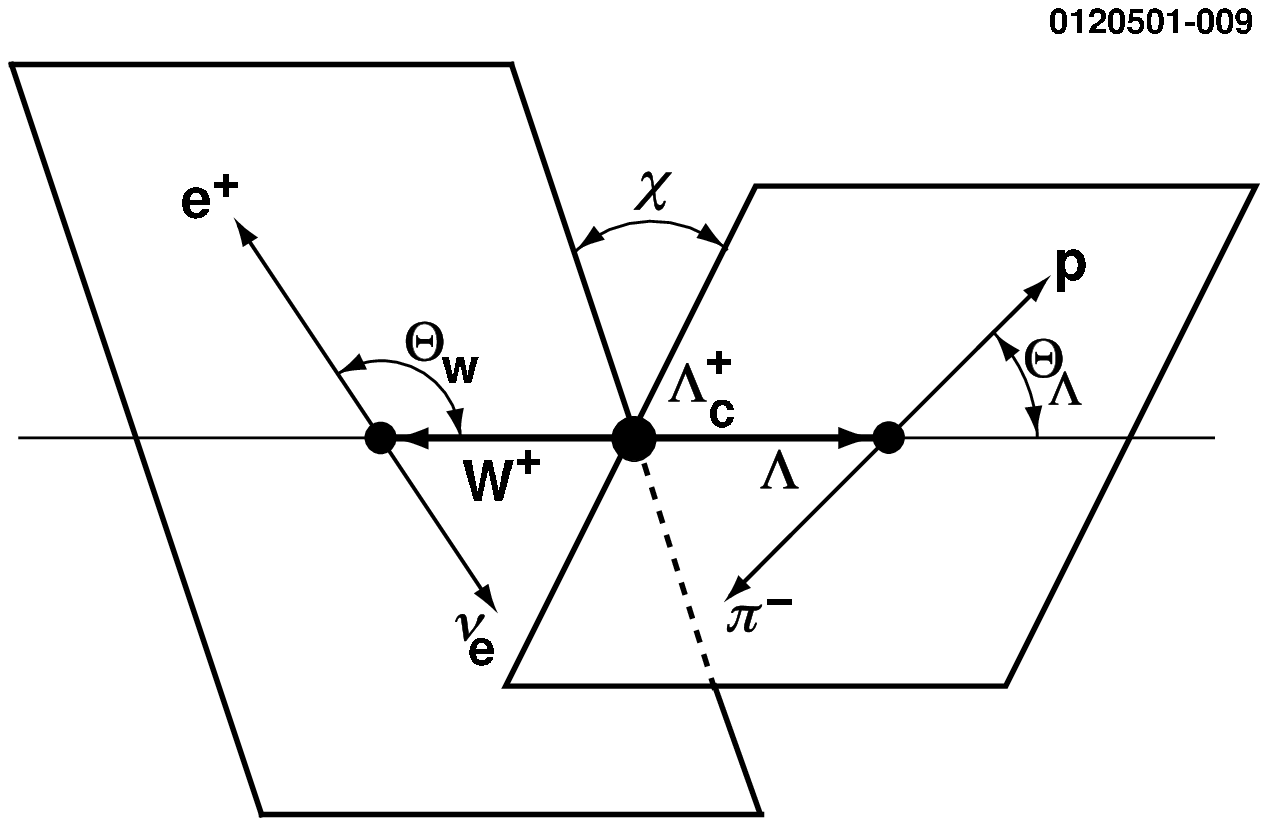}}
     \caption{\label{fig:pavlunin1} Definition of kinematic variables
              for the decay $\Lambda_c^+ \to \Lambda e^+ \nu_e$.}
  \end{minipage}
\end{figure}

This is a first publication \cite{Blusk} based on CLEO~III statistics,
4.1~fb$^{-1}$ at $\Upsilon(4S)$.
The goal is to search for the inclusive multi-body baryonic decay
$\Lambda_c^+ \to \Lambda \pi^+\pi^+ \pi^- \pi^0$.
In spectator quark model this decay is equivalent to the $D$-meson decays
observed with high rate, 
     $D^+ \to \overline{K^0}\pi^+\pi^+\pi^-\pi^0$ ($BR = 5.4^{+3.0}_{-1.4}\%$),
     $D^0 \to K^-\pi^+\pi^+\pi^-\pi^0$ ($BR = 4.4\pm0.4\%$).
In this analysis we used the power of CLEO III reconstruction
capabilities including 
$K$/$p$/$\pi$ particle identification using RICH  \& dE/dx.
We apply high momentum cut,
$ \mathbf P(\Lambda_c^+) > 3.5~GeV/c$, to suppress combinatoric background.
Other cuts were quite standard.
A clear signal from the decay
$\Lambda_c^+ \to \Lambda \pi^+\pi^+ \pi^- \pi^0$ 
is seen in the invariant $\Lambda \pi^+\pi^+ \pi^- \pi^0$ mass spectrum, 
Fig.~\ref{fig:blusk1}.

A search for substructure in $\pi^+ \pi^- \pi^0$ combinations 
shows a signal from $\Lambda_c^+ \to \Lambda \omega \pi^+$, 
$\omega \to \pi^+ \pi^- \pi^0$ and
a contribution from $\Lambda_c^+ \to \Lambda \eta \pi^+$, 
$\eta \to \pi^+ \pi^- \pi^0$
at $\sim 3\sigma$ level. The later decay has previously been observed 
in the $\eta \to \gamma \gamma$ decay mode \cite{CLEO_Ammar}.
We find that
the decay $\Lambda_c^+ \to \Lambda \pi^+\pi^+ \pi^- \pi^0$
is saturated by the $\Lambda \omega \pi^+$ and $\Lambda \eta \pi^+$ 
intermediate states.
We find NO evidence for $\omega\pi$ higher mass resonances.
This study is limited by the allowed phase space.

We use the decay $\Lambda_c^+ \to p K^- \pi^+$
for systematic cross-checks and normalization, that gives
the basic results for ratios of branchings \\
     $\frac{B(\Lambda_c^+ \to \Lambda \pi^+\pi^+ \pi^- \pi^0)}
           {B(\Lambda_c^+ \to p K^-\pi^+)} \Bigg|_{Tot.}  
           = 0.36 \pm 0.09_{stat.} \pm 0.09_{syst.} ,$     
\hfill
     $\frac{B(\Lambda_c^+ \to \Lambda \omega \pi^+ )}
           {B(\Lambda_c^+ \to p K^-\pi^+)} 
           = 0.24 \pm 0.06_{stat.} \pm 0.06_{syst.} ,$     
\\
     $\frac{B(\Lambda_c^+ \to \Lambda  \pi^+ \pi^+ \pi^- \pi^0)}
           {B(\Lambda_c^+ \to p K^-\pi^+)} \Bigg|_{N.R.}
           < 0.13~~@~~90\%~C.L.,$     
\hfill
     $\frac{B(\Lambda_c^+ \to \Lambda \eta \pi^+ )}
           {B(\Lambda_c^+ \to p K^-\pi^+)} 
           < 0.65~~@~~90\%~C.L.$

\subsection{Form Factors and Search for CP Violation in the Decay 
         $\Lambda_c^+ \to \Lambda e^+ \nu_e$}         

The Heavy Quark Effective Theory (HQET) gives a reliable prediction 
on heavy to light quark transition for $\Lambda$-type baryons using
the $V+A$ hadronic current, 
     \[<\Lambda_2|J^{V+A}_\mu|\Lambda_1> = 
     \bar u_2[f_1(q^2)\gamma_\mu(1-\gamma_5) + 
     f_2(q^2) \not{v_1} \gamma_\mu(1-\gamma_5)]u_1, \]
parameterized
in the frame of $K\ddot{o}rner$ \& $Kr\ddot{a}mer$ (KK) model \cite{KKmodel}.
Thus the 4-fold differential rate of the decay $\Lambda_c^+ \to \Lambda e^+ \nu_e$, 
     \[\frac{d\Gamma}{dq^2 d\cos \theta_{\Lambda} d\cos \theta_W d\chi} 
     \propto 
     G^2_F |V_{cs}|^2 \frac{q^2 P_{\Lambda}}{M^2_{\Lambda_c}} \cdot
     F(q^2 ,\cos \theta_{\Lambda},\cos \theta_W, \chi~|~R, M_{pole}), \]
can be expressed as a function of four kinematic variables, 
$q^2$, $\cos \theta_{\Lambda}$, $\cos \theta_W$ , $\chi$
(see Fig.~\ref{fig:pavlunin1} for explanation), 
and two parameters, 
pole mass, $M_{pole}$, 
and formfactors ratio, $R=f_2(q^2)/f_1(q^2)$, where
     $ f_i(q^2)=\frac{f_i(q^2_{max})}
        {(1-q^2/M^2_{pole})^2}(1-q^2_{max}/M^2_{pole})^2 $ is taken in dipole form.

To estimate the $\Lambda_c^+$ momentum we use three sources of information:
(i) direction of the event trust axis,
(ii) the momentum balance equation 
$\vec P_{\Lambda_c^+}^2 \simeq (\vec P_\Lambda + \vec P_e + {\vec P_\nu} )^2$,
(iii) weights from known fragmentation of $\Lambda_c^+$.

In order to extract $R$ and $M_{pole}$ we use 
4-D maximum likelihood fit \cite{Schmidt} and get
$R = -0.31 \pm 0.05_{stat} \pm 0.04_{syst}$ and
$M_{pole} = 2.13 \pm 0.07_{stat} \pm 0.10_{syst}GeV/c^2$.
Note, that $M_{pole}$ from the fit is consistent with  
$M_{D_s^{*+}}$ value used as a parameter of KK model.
The KK model predicts $R = -0.25$ 
at $M_{pole} = M_{D_s^{*+}} =2.11~GeV/c^2$.

We also measure the {\em decay asymmetry parameter}, $\alpha_{\Lambda_c^+}$,
which appears in parameterization of the differential rate after
integration over $\chi$ and $\cos\theta_W$ that gives
     $\frac{d\Gamma}{dq^2 d\cos \theta_{\Lambda}} 
      \propto
     (1 + \alpha_{\Lambda_c^+} \alpha_{\Lambda} \cos \theta_{\Lambda})$,
where $\alpha_{\Lambda}$ is a similar decay asymmetry parameter for $\Lambda$.
We obtain from the fit
     $\alpha_{\Lambda_c^+} = -0.85 \pm 0.03_{stat} \pm 0.02_{syst}$ 
     $@$ $<q^2>=0.67~(GeV/c)^2 $    
     for world average $\alpha_{\Lambda}=0.642\pm0.013$.

The decay asymmetry can be obtained for two 
     charge conjugate states separately,
     $\alpha_{\Lambda_c^+} \alpha_{\Lambda} = 
        -0.561 \pm 0.026_{stat}$,
     $\alpha_{\bar{\Lambda_c^+}} \alpha_{\bar{\Lambda}} = 
        -0.535 \pm 0.024_{stat}$.
These two values allow us to estimate a CP violation in this decay,
     $A_{\Lambda_c^+} =
            \frac{\alpha_{\Lambda_c^+} + \alpha_{\bar{\Lambda_c^+}}}
                 {\alpha_{\Lambda_c^+} - \alpha_{\bar{\Lambda_c^+}}}
     = 0.01 \pm 0.03_{stat} \pm 0.01_{syst}  \pm 0.02_{A_\Lambda}$. 
We find no evidence of CP violation in terms of the decay asymmetry parameter.


\subsection{First Search for Flavor Changing Neutral Current Decay $D^0 \to \gamma \gamma$}

Standard Model (SM) predicts the rate for the $D^0 \to \gamma \gamma$ decay
of $\sim 10^{-8}$ or less. Non SM extension, i.e.
gluino exchange in SUSY, might enhance this rate by two orders of magnitude.
A measurement of this rate is a good test of new physics beyond the SM. 

\begin{figure}[!htb]
  \begin{minipage}[t]{60mm}
    \epsfxsize=63mm
    \rightline{\epsfbox{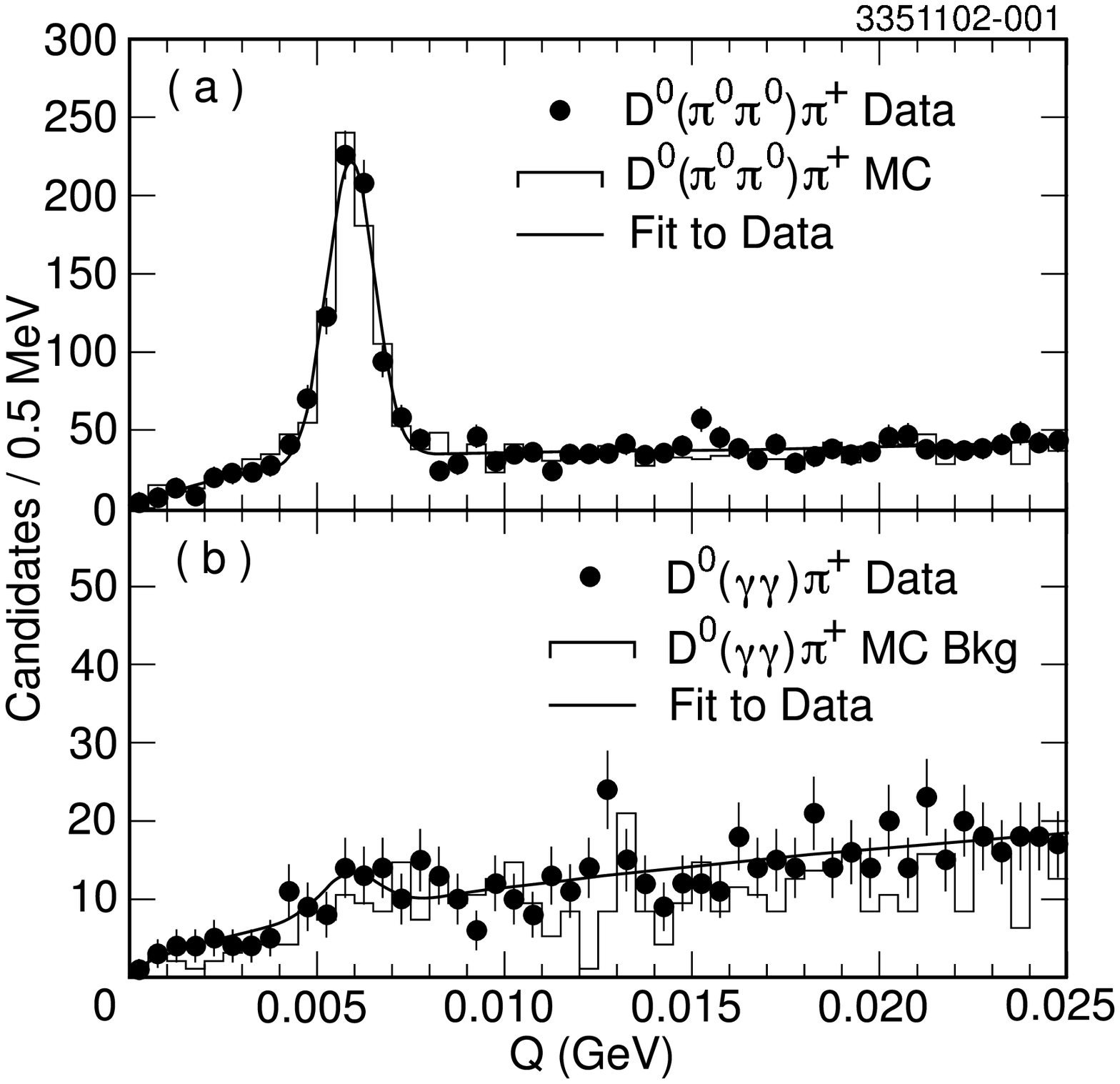}}
    \caption{\label{fig:gao1} 
             Energy release in the decay $D^{*+}\to D^0\pi+$
             for (a) $D^0\to \pi^0 \pi^0$; (b) $D^0\to \gamma \gamma$.}
  \end{minipage}
 \hfill
  \begin{minipage}[t]{95mm}
    \epsfxsize=95mm
    \centerline{\epsfbox{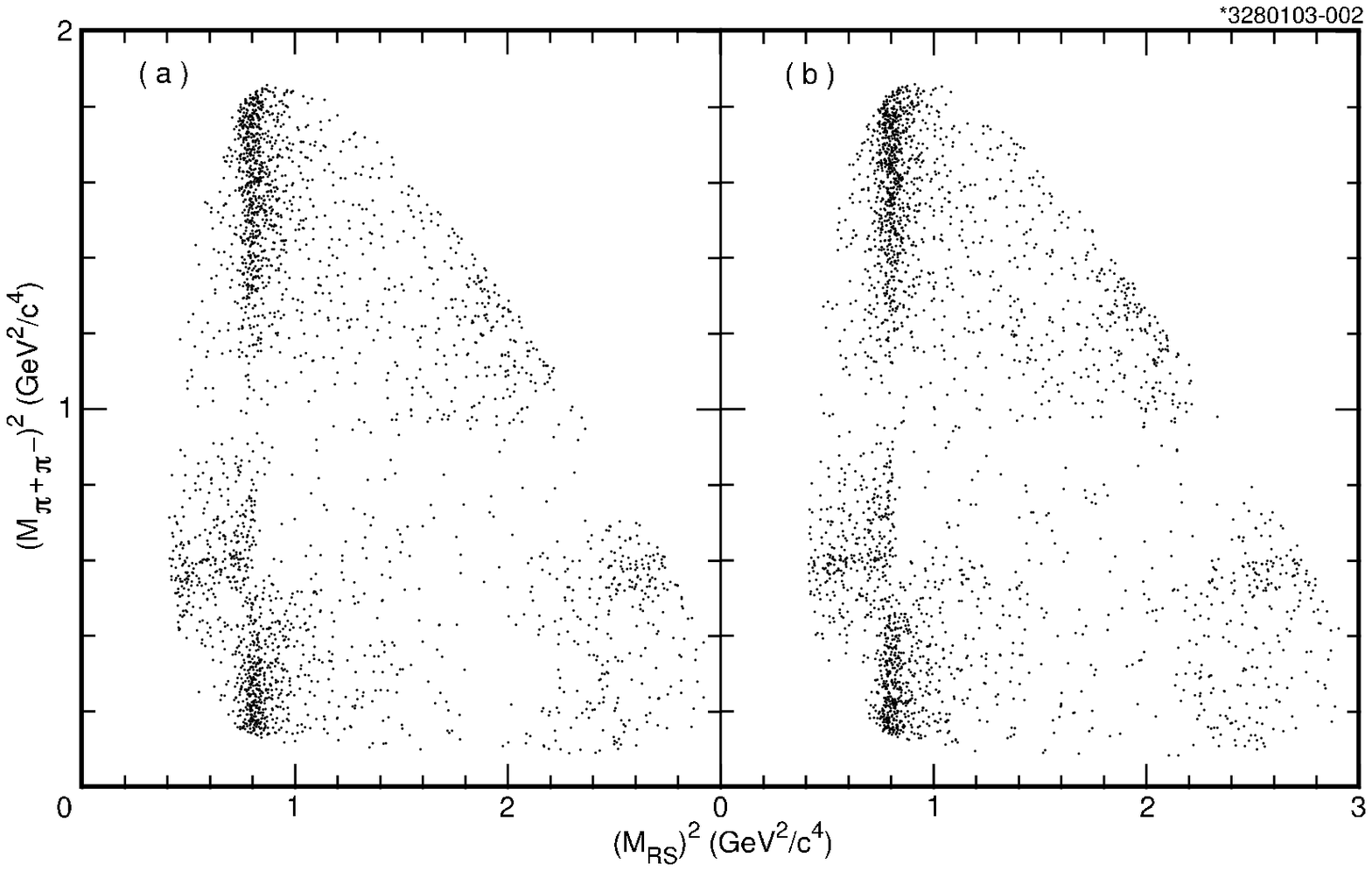}}
    \caption{\label{fig:asner1} 
     Dalitz plots for the decay $D^0 \to K^0_S \pi^+ \pi^-$ for
     (a) $D^0$,  $M_{RS}=M(K^0_S \pi^-)$;
     (b) $\overline{D^0}$, $M_{RS}=M(K^0_S \pi^+)$.
            }
  \end{minipage}
\end{figure}

For this analysis \cite{Gao} we use statistics of CLEO~II \& II.V.
Combinatoric background is suppressed by the tagging process 
$D^{*+}\to D^0\pi^+$
using the energy release variable $Q=M(D^{*+})-M(D^0)-m_{\pi^+}$,
Fig.~\ref{fig:gao1}b. 
The decay $D^0\to \pi^0 \pi^0$ with measured rate 
is used for systematic cross-checks and normalization, Fig.~\ref{fig:gao1}a.
The ratio of efficiency is determined from Monte Carlo.
$\varepsilon(\gamma \gamma)/\varepsilon(\pi^0 \pi^0) = 1.59\pm 0.05$.
We found $19.2\pm9.3$ signal events from the fit, 
and set an upper limit 
     \[ B(D^0\to \gamma \gamma) / B(D^0\to \pi^0 \pi^0) < 0.033, 
   ~~~B(D^0\to \gamma \gamma) <2.9\times10^{-5}~~@~~90\% C.L. \]

\subsection{Dalitz Plot Analyses of the Decays $D \to PPP$ at CLEO }

The Dalitz plot analyses of $D$-meson decays in to three pseudoscalars,
$D \to PPP$, are interesting for many reasons.
\begin{itemize}
\vspace{-3mm} \item 
     The Dalitz plot clearly shows the intermediate states and their interference.
\vspace{-3mm} \item 
     These sorts of decays accounts for a large fraction of the total
D meson decay rate.
\vspace{-3mm} \item 
     It is a good place to study scalars, $f_0(980)$, $a_0(980)$.
\vspace{-3mm} \item 
     In these decays there are indications
     on the not well established light scalar mesons,
     $\sigma(500)$, $\kappa(400-600)$
     (see E791, FOCUS, CLEO recent results).
\vspace{-3mm} \item 
     An opportunity to search for CP violation 
     in amplitudes and phases, that may be more visible than rate  
     asymmetries.
\end{itemize}
\vspace{-3mm}
     For the Dalitz plot analyses discussed here we use
     the CLEO II.V 9~fb$^{-1}$ data sample, the large sample 
     which has well understood resolutions and systematics.
     We also use a ``generic'' (all known decays included) 
     and signal Monte-Carlo simulation in order to estimate an efficiency
     variation across the Dalitz plot.

     We exploit $D^{*} \to D \pi_{slow}$ tagging process in order 
     to suppress background and
     to define an initial flavor of {$D^0$ or $\overline{D^0}$}-meson.
     We use other standard for CLEO event selection algorithms, and
     $\sim 2-3\sigma$ cuts on energy release variable, $Q=M(D^*)-M(D)-M_{\pi}$, and 
     invariant $D$-meson mass, $M(D)$.

      The Dalitz plot is a scatter plot of two (of three) 
      invariant masses squared, i.e. $m^2_{23}$  versus $m^2_{12}$.
      A general formalism of the 
      {\em matrix element parameterization} and 
      {\em unbinned maximum likelihood fit}
      is published in detail in Ref.~\cite{Kopp}.
Thus for the matrix element we use a sum of amplitudes 
of non-resonant and all possible resonant terms,
\[
       {\mathcal M} = a_{nr}e^{i\varphi_{nr}} + 
        \sum_{resonances} a_{r}e^{i\varphi_{r}} {\mathcal A}_J(123|r),
\]
where $a_r e^{i\varphi_r}$ are the complex parameters (fit parameters)
and ${\mathcal A}_J(123|r)$ is a spin dependent Breit -Wigner amplitude
for appropriate intermediate resonance, $r$. 

Based on parameters extracted from the fit one may define 
for each resonance a 
${Fit~Fraction} = \frac{\oint_{PS}|a_r {\mathcal A}_J(123|r)|^2 d{\mathcal DP}}
{\oint_{PS}|{\mathcal M}|^2 d{\mathcal DP}}$,
and an {\em Integrated CP Asymmetry},
${\mathcal A}_{CP} = \oint_{PS} 
\frac{ |{\mathcal M}_{D^0}|^2 - |{\mathcal M}_{\bar{D^0}}|^2 }
     { |{\mathcal M}_{D^0}|^2 + |{\mathcal M}_{\bar{D^0}}|^2 } d{\mathcal DP} \bigg/
      {\oint_{PS} d{\mathcal DP} }.$
The modified complex parameters can be also used in order to search for
CP violation in terms of amplitudes and phases,
     $a_r e^{i\varphi_r} \longrightarrow 
      a_r e^{i\delta_r}\bigg(1 \pm \frac{b_r}{a_r} e^{\pm i\phi_r} \bigg),$
     where 
     $a_r$ and $\delta_r$ - CP conserving parameters; 
     $b_r$ and $\phi_r$ - CP violating parameters, 
     showing the difference in amplitude and phase between $D^0$ and $\overline{D^0}$.

         
\subsection{Search for CP Violation in the Decay $D^0 \to K^0_S \pi^+ \pi^-$}

The results of the 
Dalitz plot analysis of $D^0 \to K^0_S \pi^+ \pi^-$ were 
published \cite{asner1}
for clearly observed and measured 10 modes,
     $K^{*-}(892)\pi^+$,
     $K_0^*(1430)^-\pi^+$,
     $K_2^*(1430)^-\pi^+$, 
     $K^*(1680)^-\pi^+$,
     $K_S^0\rho$,
     $K_S^0\omega$, 
     $K_S^0 f_0(980)$,
     $K_S^0 f_2(1270)$,
     $K_S^0 f_0(1370)$, 
     and the ``wrong sign'' 
     $K^{*+}(892)\pi^-$.

In present consideration we use the same clean sample of 5299$\pm$73 signal 
events with $\sim 2\%$ of background
to search for CP violation  (see  Fig.~\ref{fig:asner1}).
The SM predicts for this decay a CP violation in rate at the level of $\sim10^{-6}$ 
due to the $K^0-\overline{K^0}$ mixing.
Hence, an observation of CP violation in $D^0 \to K^0_S \pi^+ \pi^-$ 
would be a strong evidence for non-SM processes.



     The results will be published soon
     in terms of parameters,
     $a_r$, $\delta_r$,
     $b_r/a_r$, $\phi_r$, 
     {\em Fit Fraction} and 
     {\em Interference Contribution}  
     for 10 intermediate states 
     and non-resonant amplitude.
Preliminarily, we find NO indication to CP asymmetry in integrated parameter,
${\mathcal A}_{CP} = -0.39\pm 0.34$, and in all other types
of CP asymmetries.


\subsection{Dalitz Plot Analysis of the Decay $D^0 \to \pi^+ \pi^- \pi^0$}

The decay $D^0 \to \pi^+ \pi^- \pi^0$ is interesting for a couple of reasons:
\begin{itemize}
\vspace{-3mm} \item
     SM predicts a large CP violation in rate $\sim10^{-3}$
     that is close to our sensitivity; 
\vspace{-3mm} \item
     E791 found in $D^+ \to \pi^+ \pi^+ \pi^-$ decay
     strong evidence for $\sigma(500)\to\pi^+ \pi^-$ in the intermediate state.
     We would expect to observe this state or its charged isospin partner in 
     $D^0 \to \pi^+ \pi^- \pi^0$.
\end{itemize}

\begin{figure}[!htb]
  \begin{minipage}[t]{50mm}
    \epsfxsize=60mm    
    \centerline{\epsfbox{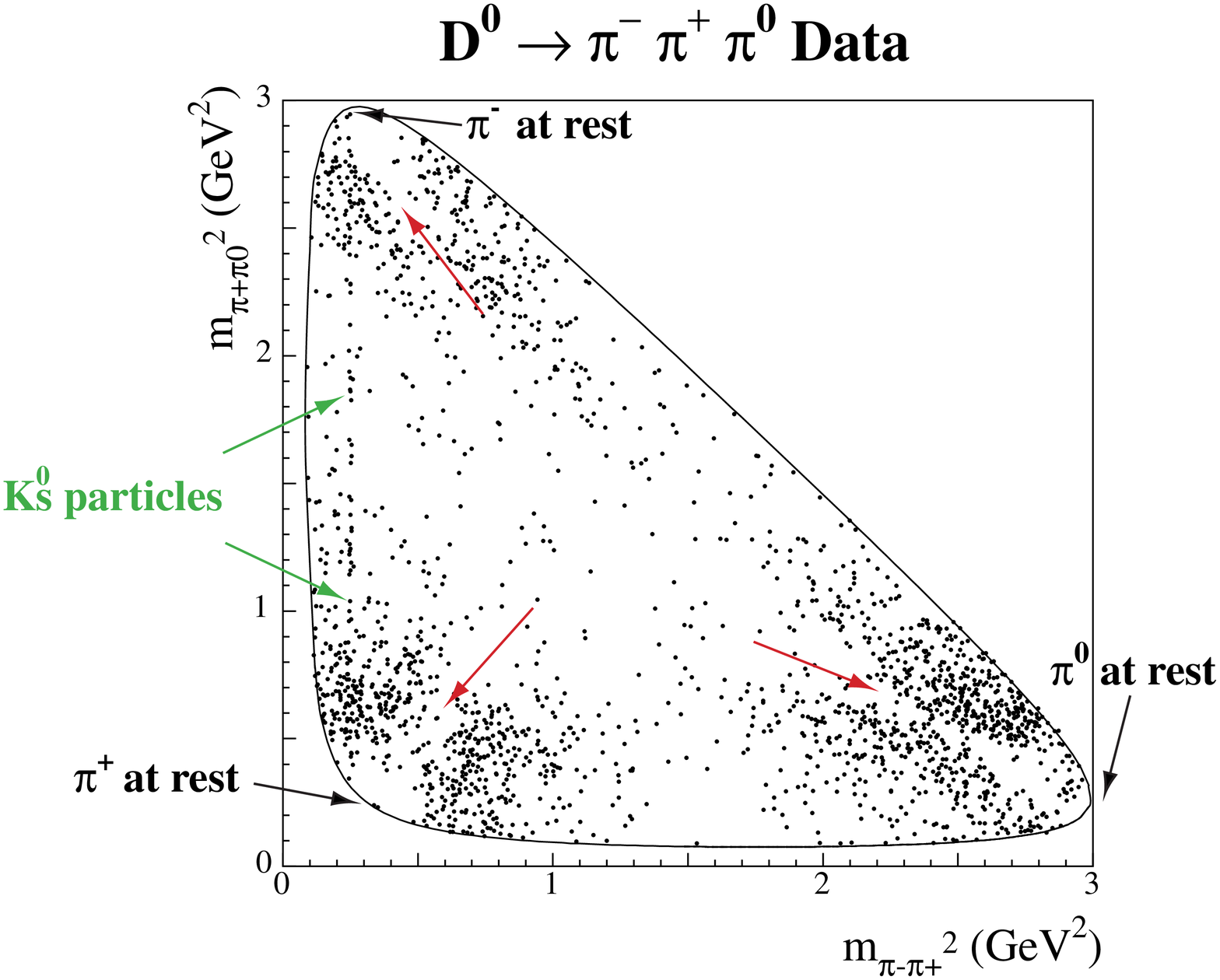}}
    \caption{\label{fig:plager1} The Dalitz plot for $D^0 \to \pi^+ \pi^- \pi^0$ decay.}
  \end{minipage}
\hfill
  \begin{minipage}[t]{32mm}
    \epsfxsize=35mm
    \centerline{\epsfbox{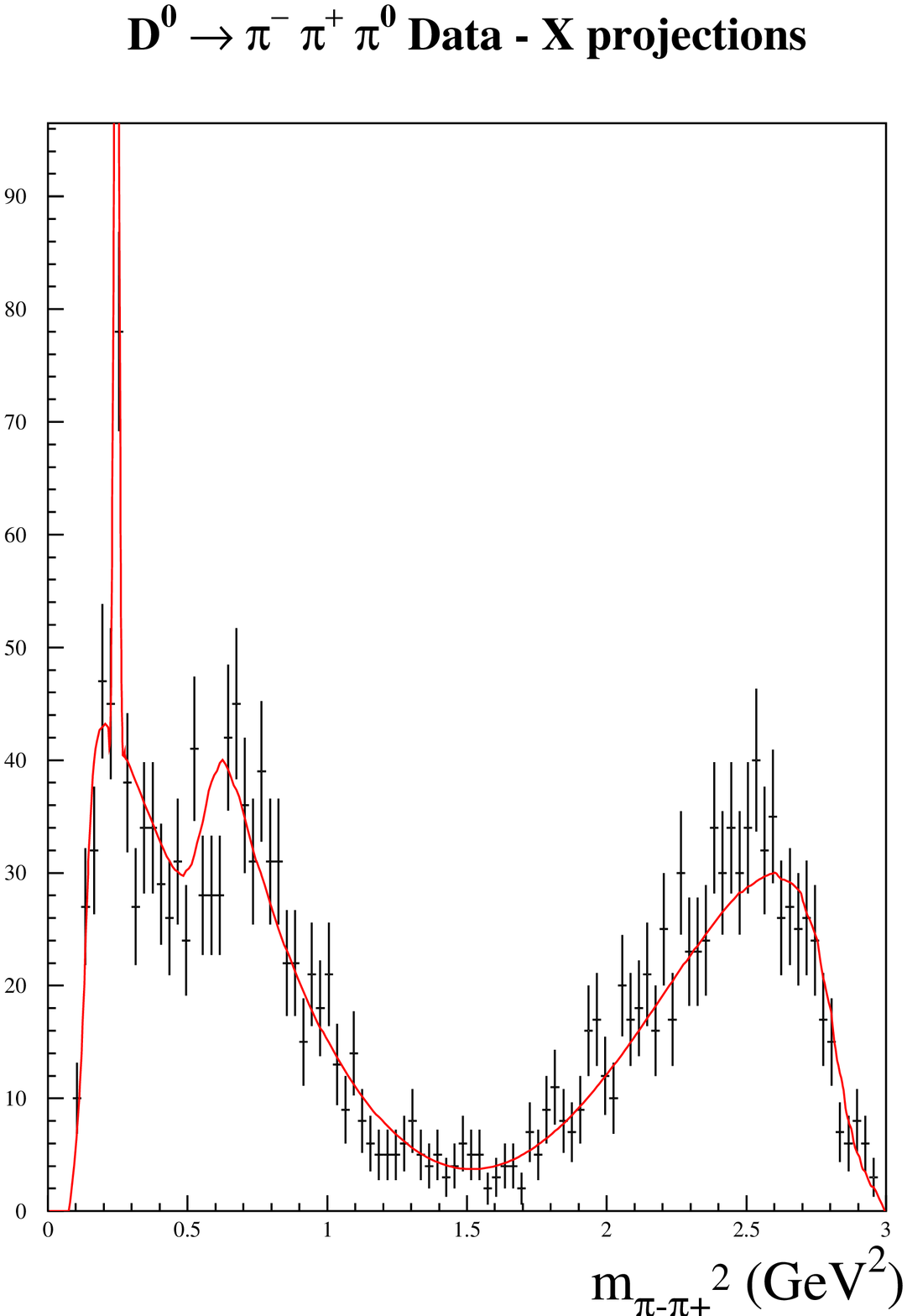}}
    \caption{\label{fig:plager2} The $m^2_{\pi^-\pi^+}$ projection.}
  \end{minipage}
\hfill
  \begin{minipage}[t]{32mm}
    \epsfxsize=35mm
    \centerline{\epsfbox{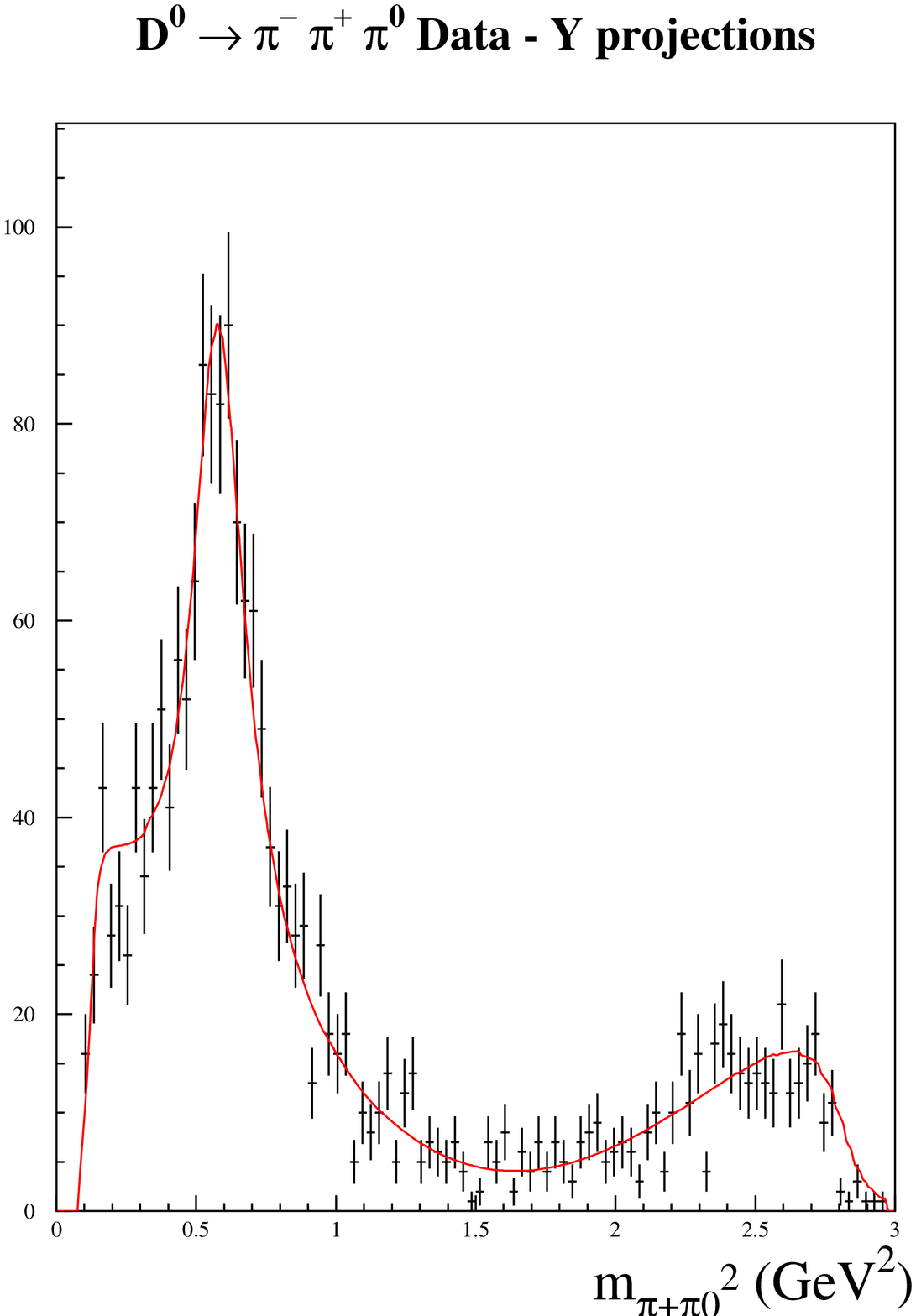}}
    \caption{\label{fig:plager3} The $m^2_{\pi^+\pi^0}$ projection.}
  \end{minipage}
\hfill
  \begin{minipage}[t]{32mm}
    \epsfxsize=35mm
    \centerline{\epsfbox{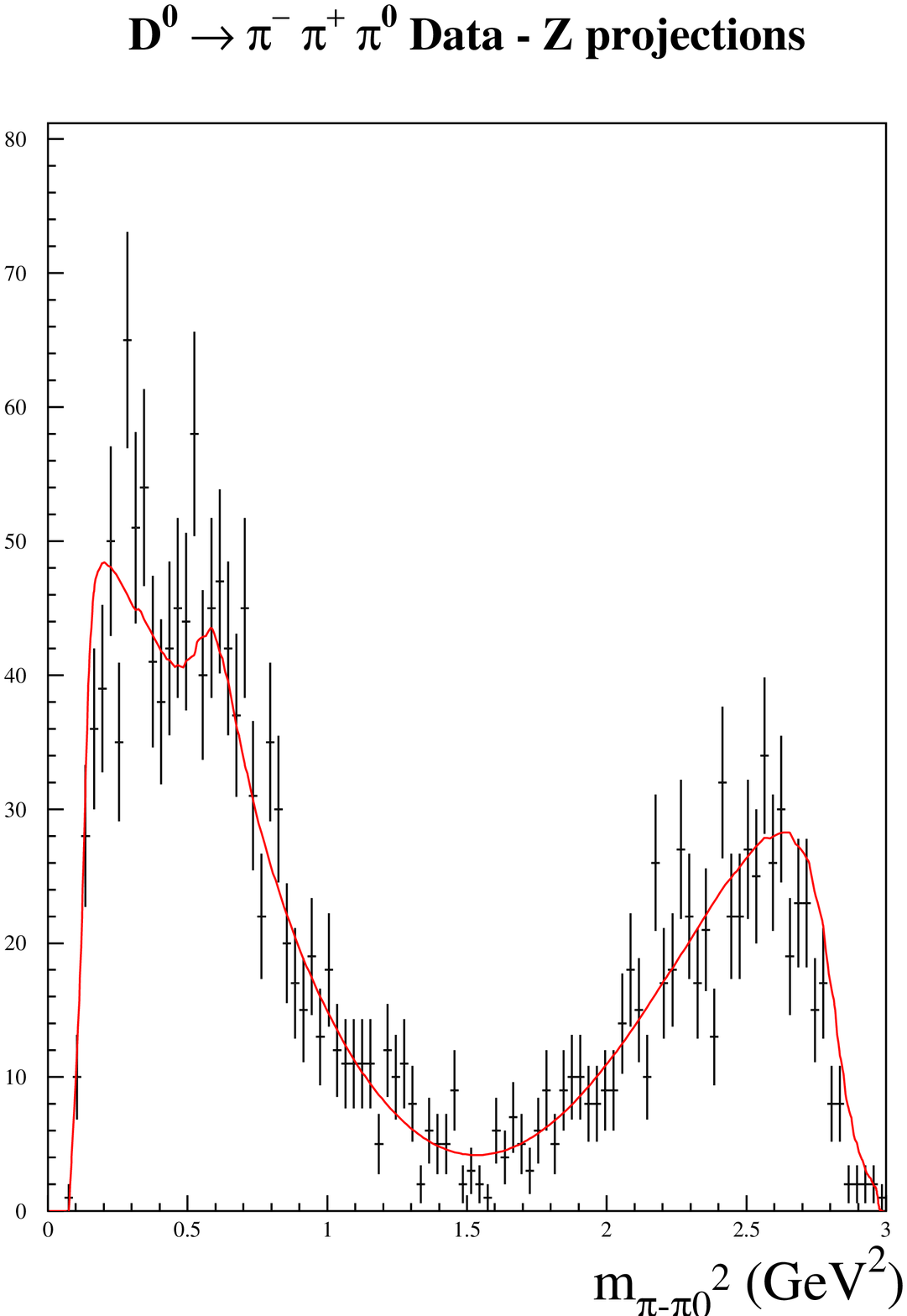}}
    \caption{\label{fig:plager4} The $m^2_{\pi^-\pi^0}$ projection.}
  \end{minipage}
\end{figure}

\begin{table}[!htb]
\begin{center}
\caption{\label{tab:plager1} 
         Results of the Fit for the decay $D^0 \to \pi^+ \pi^- \pi^0$.
         Statistical and systematic errors are shown respectively.}
\begin{tabular}{|c|c|c|c|}
\hline
Interm. state    & Amplitude              & Phase ($^o$)     & Fit Fraction (\%)\\
\hline
$\rho^+ \pi^-$   & 1 (fixed)              & 0 (fixed)        & 76.5$\pm$1.8$\pm$4.8 \\
$\rho^0 \pi^0$   & 0.56$\pm$0.02$\pm$0.07 &  10$\pm$3$\pm$3  & 23.9$\pm$1.8$\pm$4.6 \\ 
$\rho^- \pi^+$   & 0.65$\pm$0.03$\pm$0.04 &  -4$\pm$3$\pm$4  & 32.3$\pm$2.1$\pm$2.2 \\ 
Non resonant     & 1.03$\pm$0.17$\pm$0.31 &  77$\pm$8$\pm$11 &  2.7$\pm$0.9$\pm$1.7 \\  
\hline
\end{tabular}
\end{center}
\end{table}

     Distributions of $Q$ and $M(D)$ shows $\sim 1100$ signal
     events with $\sim 18\pm 3\%$ of background. 
     The Dalitz plot and three projections are shown in 
     Figs~\ref{fig:plager1}--\ref{fig:plager4}.
The observed $K^0_S \pi^0$ contribution is consistent with previous measurements
and is not allowed to interferee with the other contributions to the Dalitz plot.
Preliminary results of the fit parameters are listed in Table~\ref{tab:plager1}. 
Our conclusions:
$ A_{CP} = 0.01^{+0.09}_{-0.07}(stat)\pm0.09(syst)$ and
there is NO CP asymmetry {in $A_{CP}$} at the level of our sensitivity;
there is NO statistically significant difference {\em in amplitude and phase} between 
$D^0/\overline{D^0} \to \pi^+ \pi^- \pi^0$;
non-resonant contribution is small;
there is NO indication on $\sigma(500)$ fit fraction at the level of $\sim$1\%.


\subsection{First Observation and Dalitz Plot Analysis of the Decay $D^0 \to K^0_S \eta \pi^0$}

     In the decay of $D^0 \to K^0_S \eta \pi^0$ one would expect 
     a strong manifestation of $D^0 \to K^0_S a_0(980)$ ,
     if the scalar resonance $a_0(980)$ indeed contributes 
     to the complementary $D^0 \to K^0_S (K^+ K^-)$
     decay along with $f_0(980)$.
     There is no result in PDG for $D^0 \to K^0_S \eta \pi^0$ yet.

\begin{figure}[!htb]
  \begin{minipage}[t]{50mm}
    \epsfxsize=50mm
    \centerline{\epsfbox{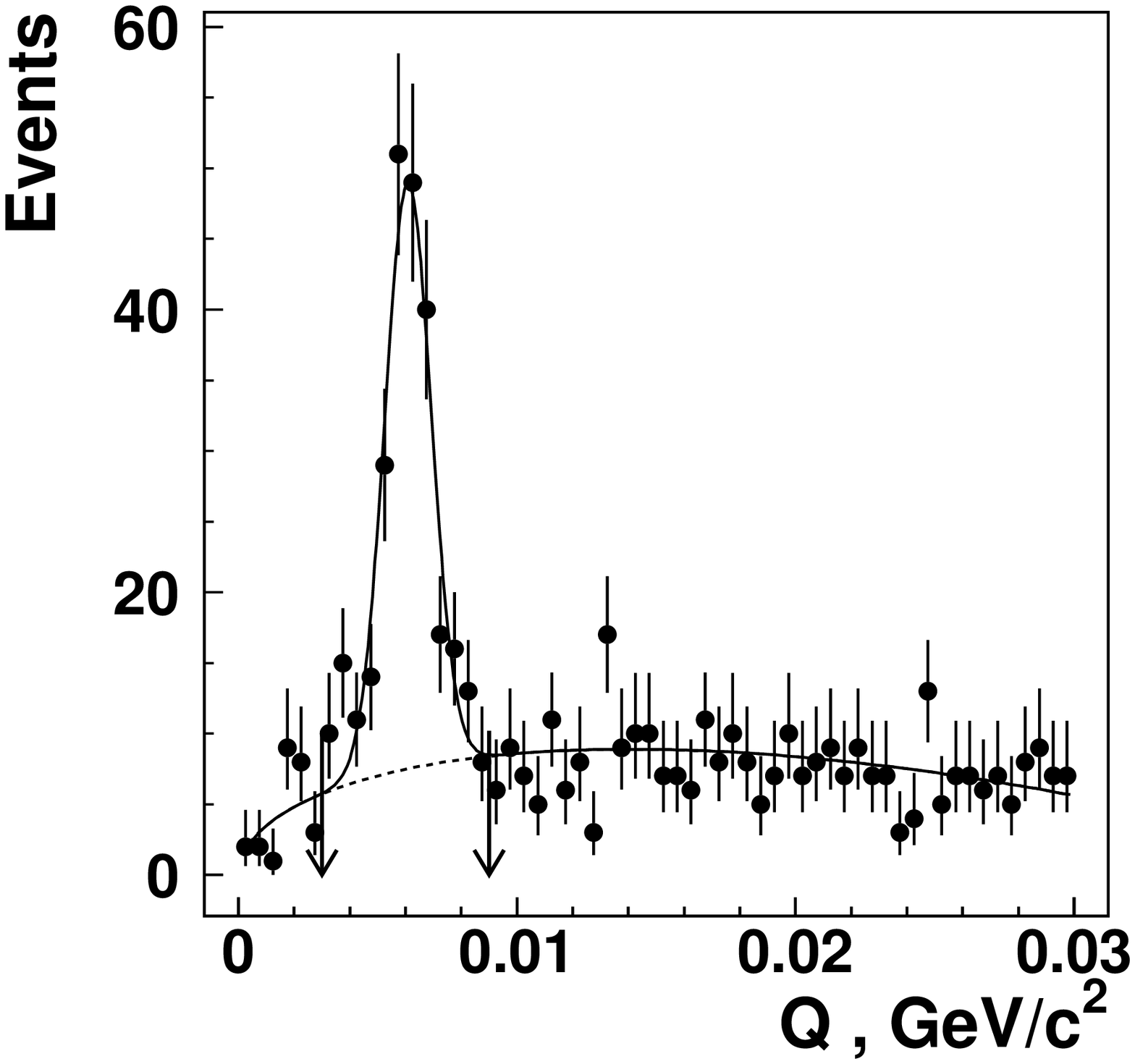}}
    \caption{\label{fig:dubrovin1} The Q distribution for the decay
            $D^{*+} \to D^0(K^0_S \eta \pi^0) \pi^+$.}
  \end{minipage}
 \hfill 
  \begin{minipage}[t]{50mm}
    \epsfxsize=50mm
    \centerline{\epsfbox{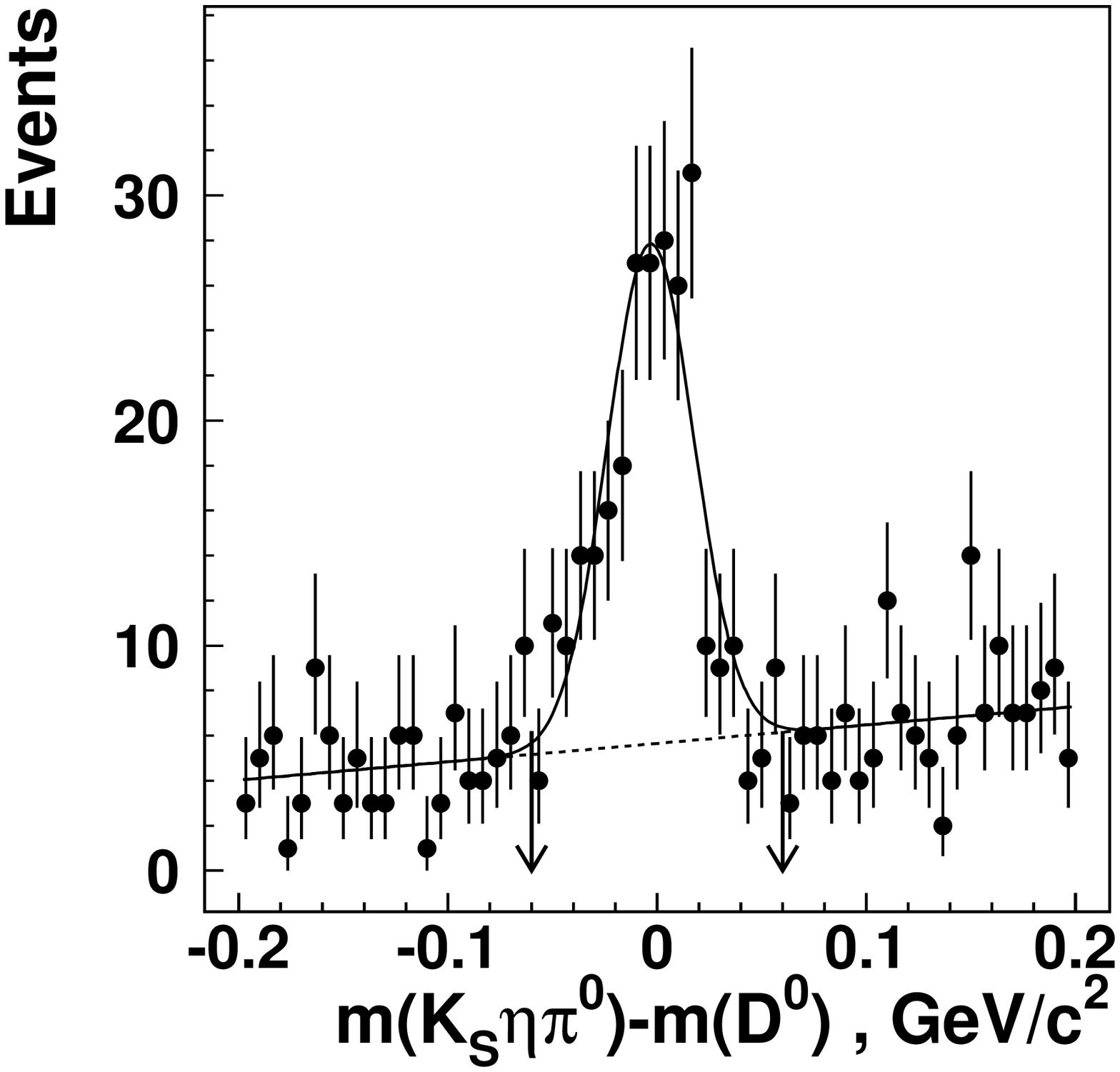}}
    \caption{\label{fig:dubrovin2} 
     The $M(K^0_S \eta \pi^0) - M_{D^0}$ mass difference distribution.}
  \end{minipage}
 \hfill 
  \begin{minipage}[t]{50mm}
    \epsfxsize=50mm
    \rightline{\epsfbox{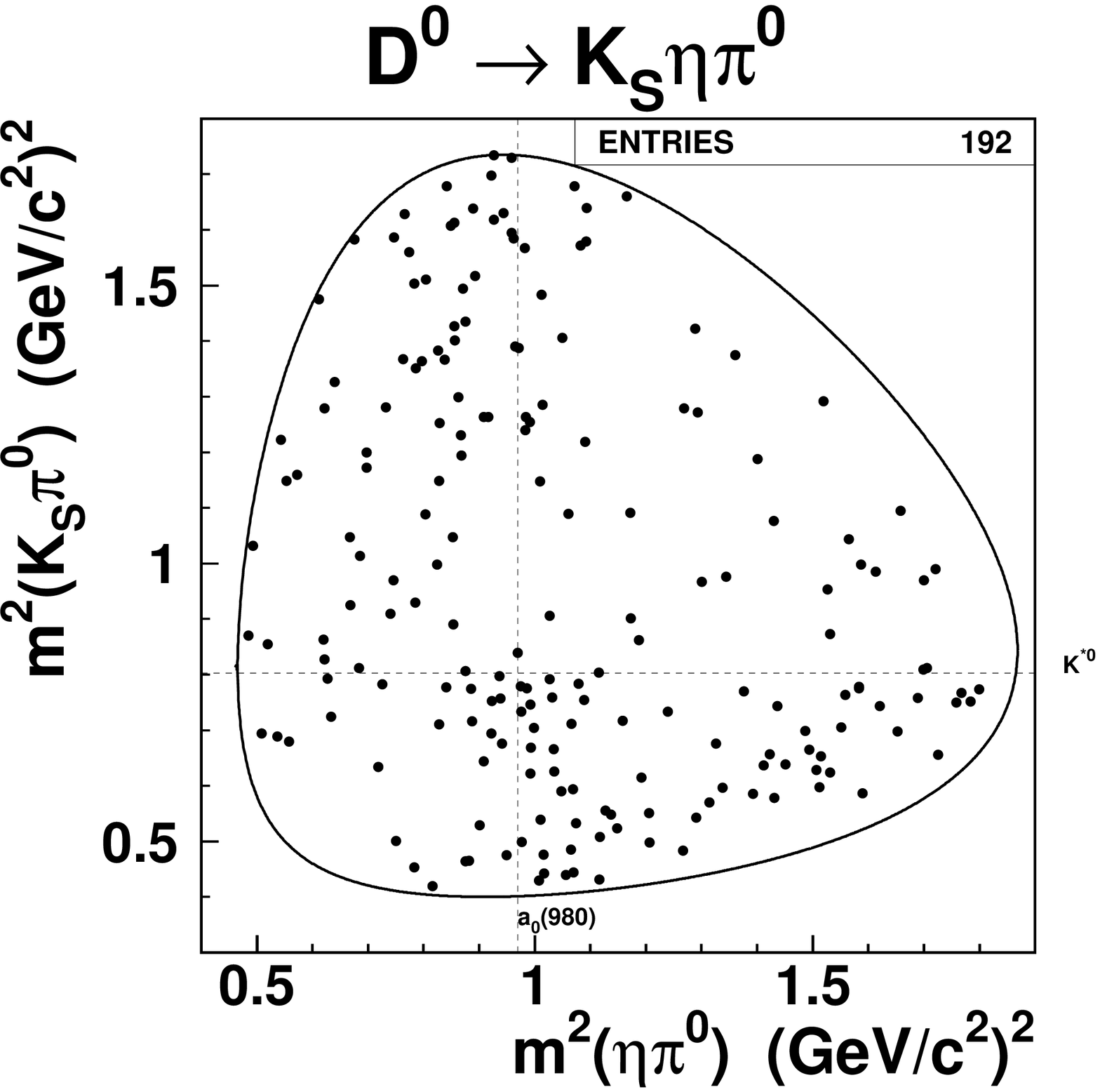}}
    \caption{\label{fig:dubrovin3} Dalitz plot for the decay $D^0 \to K^0_S \eta \pi^0$.}
  \end{minipage}
\end{figure}

     To study this decay we use the mode: 
$K^0_S \to \pi^+\pi^-$, $\eta \to \gamma \gamma$, $\pi^0 \to \gamma \gamma$.
The decay 
$D^0 \to K^0_S \pi^0$ 
is used for systematic cross-checks and normalization.
We observe a clean signal in $Q$ (Fig.~\ref{fig:dubrovin1}) 
and $M(K^0_S \eta \pi^0)$ (Fig.~\ref{fig:dubrovin2})
distributions and have preliminary measured the ratio of branchings
\[ 
     R = \frac{BR(D^0 \to K^0_S \eta \pi^0)} {BR(D^0 \to K^0_S \pi^0)}
       = 0.38 \pm 0.07_{stat.} \pm 0.05_{syst.}.
\]

In spite of large background $\sim 30\%$ we are going to perform a 
Dalitz plot analysis for the decay $D^0 \to K^0_S \eta \pi^0$
(see Fig.~\ref{fig:dubrovin3}). Our preliminary observations
confirms that the $K^0_S a_0(980)$ is a dominant intermediate state, 
the $K^*(892)\eta$ is seen with a significant interference term.
To get a good fit to the observed data another contribution, such as
non-resonant or high mass states, is needed.
We are working on completing this analysis.



\section{CLEO-c and CESR-c project news}

The NSF has approved a five year program of 
charm physics called CLEO-c and CESR-c \cite{CLEOc-CESRc}.
The first year of this program has been successfully completed.
Since October 2002 to March 2003 the crucial part of this project
a single super-conducting (SC) wiggler had been successfully tested at CESR
and five other wigglers have been assembled and prepared for installation.
During a current shutdown since March 3 to July 2, 2003 the 6 SC wigglers
will be installed in CESR. 
CLEO III will be upgraded to CLEO-c by replacing 
the silicon vertex detector (SVX) with a 6-layers drift chamber 
with large stereo angle (ZD). 

The program of CLEO-c scans can be briefly listed as follow:
\begin{itemize}
\vspace{-2mm} \item
     2003 Act I: $\psi (3770)$ --- 3~fb$^{-1}$; 
     30M events, 6M {\em tagged} D
     (310 times MARK III).
\vspace{-2mm} \item 
     2004 Act II: $\sqrt{s} \sim 4100~MeV$ --- 3~fb$^{-1}$; 
     1.5M $D_s \bar D_s$, 0.3M {\em tagged} $D_s$  
     (480 $\times$ MARK III).
\vspace{-2mm} \item 
     2005 Act III: $\psi (3100)$ --- 1~fb$^{-1}$; 
     1~Billion $J/\psi$ 
     (170 times MARK III, 20 times BES II).
\end{itemize}
This statistics is required for precise measurement of branching ratios,
decay constants and other SM parameters in charm sector.

\section{Summary and Acknowledgments}
     The CLEO Collaboration continues to produce results in charm physics 
using CLEO~II, II.V, III statistics.
We have a broad charm-physics dedicated program and  
are looking forward to CLEO-c/CESR-c data.
We gratefully acknowledge the effort of the CESR staff
in providing us with
excellent luminosity and running conditions.
This work was supported by 
the National Science Foundation,
the U.S. Department of Energy,
the Research Corporation,
and the 
Texas Advanced Research Program.


\end{document}